\documentclass[preprint,showpacs,preprintnumbers,amsmath,amssymb,nofootinbib]{revtex4}

\usepackage{epsfig}

\newcommand{\be}{\begin{equation}}
\newcommand{\ee}{\end{equation}}
\newcommand{\beqq}{\setlength\arraycolsep{2pt}\begin{eqnarray}}
\newcommand{\eeqq}{\vspace{0cm} \end{eqnarray}}
\newcommand{\bea}{\begin{eqnarray}}
\newcommand{\eea}{\end{eqnarray}}

\begin{document}

\title{Quantized fields and gravitational particle creation in $f(R)$ expanding universes}

\author{S. H. Pereira$^1$}\email{saulopereira@unifei.edu.br}
\author{C. H. G. Bessa$^2$}\email{chgbessa@astro.iag.usp.br}
\author{J. A. S. Lima$^{2,3}$}\email{limajas@astro.iag.usp.br}

\vspace{0.5cm}

\affiliation{$^1$Universidade Federal de Itajub\'a, Campus Itabira \\ Rua S\~ao Paulo, 377 -- 35900-373, Itabira, MG, Brazil}
\affiliation{$^2$Departamento de Astronomia (IAGUSP),  Universidade de S\~{a}o
Paulo \\ Rua do Mat\~ao, 1226 -- 05508-900, S\~ao Paulo, SP, Brazil}
\affiliation{$^{3}$Center for Cosmology and Astro-Particle Physics, The Ohio State University, \\  
191 West Woodruff Avenue, Columbus, OH 43210, USA}

\vspace{1.0cm}

\begin{abstract}
The problem of  cosmological particle creation for a spatially flat, homogeneous and isotropic
Universes is discussed in the context of $f(R)$ theories of gravity. 
Different from cosmological models based on general relativity theory, it is found that a conformal invariant metric does not forbid the 
creation of massless particles during the early stages (radiation era) of the Universe. 
\end{abstract}


\maketitle
\section{Introduction}

It is widely believed that the possible emergence of space and time trough a cosmic singularity and the presence of horizons in the Friedman-Robertson-Walker (FRW) models strongly suggest that matter and radiation need to be somehow created in  order to overcome some basic conceptual difficulties of big-bang cosmology \cite{GWS83}. In these connections, the process of  matter creation in an expanding universe has been extensively discussed in the last four decades either from a macroscopic, as well as from  microscopic viewpoints [2-18].
 
Macroscopically, the matter creation was extensively investigated as a byproduct of bulk viscosity processes near the Planck era as well as in the slow-rollover phase of the new inflationary scenario \cite{Zeld70,Murphy73,Hu82,Barrow86}. Later on,  the first 
self-consistent macroscopic formulation of the matter creation process 
was put forward by Prigogine and coworkers \cite{Prigogine} and formulated in a covariant way
by Calv\~{a}o {\it et al.} \cite{CLW} with basis on the relativistic nonequilibrium thermodynamics.  
In comparison to the standard equilibrium 
equations, the process of creation  at the expense of the gravitational field is described by two new ingredients: a balance 
equation for the particle  number density and a negative  
pressure term in the stress tensor. Such quantities 
are related to each  other in a very definite way by the second law of 
thermodynamics. In  particular, the creation pressure depends on the creation rate and may 
operate, at level of Einstein's equations, to prevent either a spacetime 
singularity \cite{Prigogine,AGL} or to generate an early inflationary 
phase \cite{LGA}.  More recently, such formulation has also been applied to explain the late time accelerating stage of the Universe and other complementary observations \cite{zimd,LSS08}.

Microscopically,  after the pioneering work of Parker \cite{park68}, the quantum process of particle creation  
in the course of the cosmological expansion has also been studied by several authors \cite{park69,books,mukh,partcrea,grav,gribmama}.  These `Parker Particles' are created because, according to the covariant equations of the fields in the Heisenberg picture, the positive and negative frequency parts of the fields become mixed during the universe expansion, so that the creation and annihilation operators at one time $t_1$ are linear combinations of those ones at an earlier time $t_2$, resulting at first, in a particle production. 

One of the most interesting results from Parker's work is that in a radiation dominated Universe, the positive- and negative- frequency mode functions for massless fields are not mixed in the course of the expansion due to the conformal invariance of the metric. In particular, this means that there is no creation of massless particles, either of zero or non-zero spin. Viewed in another way, the dispersion relation in Fourier space  is exactly the same one of the flat Minkowski space, and, as such,  a wave propagation of a massless particle is not accompanied by particle production. As a consequence, in the GR framework, no photon, graviton or any other kind of massless particles are produced by purely expanding effects at early times \cite{grav}. 

Usually, the calculations of particle production  deal with comparing the particle number at asymptotically early and late times, or with respect to the vacuum states defined in two different frames and do not involve any loop calculation. However, the main problem one encounters when treating quantization in expanding backgrounds concerns the interpretation of the field theory in terms of particles. The absence of Poincar\'e group symmetry in curved space-time leads to the problem of the definition of particles and vacuum states. The problem  may be solved by using the method of the diagonalization of instantaneous Hamiltonian  by a Bogoliubov transformation, which leads to finite results for the number of created particles \cite{gribmama}.

On the other hand, the major developments concerning the study of particle production in curved background are concentrated in the standard general relativity. Recently, non-standard gravity theories have been proposed as an alternative to explain the present accelerating stage of the universe only with cold dark matter, that is, with no appealing to the existence of dark energy. This reduction of the so-called dark sector is naturally obtained in $f(R)$ gravity theories. In such an approach, the curvature scalar  $R$ in the Einstein-Hilbert action is replaced by a general function $f(R)$, so that the Einstein field equation is recovered as a particular case \cite{fR,allemandi}. 

A basic  motivation for this kind of generalization is the fact that higher order terms in curvature invariants have to be added to the effective Lagrangian of the gravitational field when quantum corrections are taken into account \cite{gaspe,vilk,staro}. Two different formalisms are usually adopted to study $f(R)$ theories. In the so-called metric approach, the equations of motion are obtained by varying the metric tensor, while in the Palatini formalism, one considers the metric and the affine connection to be independent of each other, and one has to vary the action with respect to both of them. These two methods lead to the same equations only if $f(R)$ is a linear function of $R$ (GR case). The metric approach leads to fourth-order equations driving the evolution of the scale factor while the Palatini method leads to second-order equations, so that it is appealing because of its simplicity. Actually, for many interesting examples, it has been argued that some problems of instability are present in the metric approach \cite{dolgov}, although some recent works have cast doubts on the existence of such instabilities \cite{cembranos}. In what follows, we will focus our attention on the Palatini formalism and a simple power law type of $f(R)$. As shown by Vollick \cite{fR} using the Palatine approach,  the inclusion  of a  $R^{-1}$ term in the action leads to a theory of gravity that predicts late time accelerated expansion of the Universe. Sotiriou \cite{sotiriou} also demonstrated  that a $R^3$ term can account for an early time inflation, contrary to a $R^2$ type discussed by many authors (see Meng and Wang \cite{meng} and Refs. there in).  More recently, Miranda {\it et al.} discussed  a viable $f(R)$ cosmology based  on a logarithm contribution of the scale factor which emerges naturally as a limiting case of a general power law \cite{waga09}. 

In this letter we will present only the basic ideas and results on the scalar particle creation on expanding universe in the view of a $R^{n}$ type $f(R)$ theory. In brief, our basic result can be stated as follows:  massless particles in a radiation dominated Universe can be produced by purely expanding effects in generic $f(R)$ theories. In other words, Parker's result is valid only in the context of general relativity. 

\section{Quantization in Expanding Backgrounds}

The canonical quantization of a scalar field in  curved backgrounds follows in straight analogy with the quantization in a flat Minkowski background with the gravitational metric treated as a classical external field which is generally nonhomogeneous and nonstationary. Let us summarize the basic results for a linear Klein-Gordon scalar field in a spatially flat FRW geometry (in our units $c=1$).  

A real minimally coupled massive scalar field $\phi$ is described in a curved spacetime by the action \cite{books,mukh}
\begin{equation}\label{m63}
S={1\over 2} \int \sqrt{-g}d^4 x \bigg[g^{\alpha\beta}\partial_\alpha\phi\partial_\beta\phi-m^2 \phi^2\bigg]\,.
\end{equation}
In terms of the conformal time $\eta$, the metric tensor $g_{\mu\nu}$ is conformally equivalent to the Minkowski metric $\eta_{\mu\nu}$, so that the line element is  $ds^2=a^2(\eta)\eta_{\mu\nu}dx^\mu dx^\nu$, where $a(\eta)$ is the cosmological scale factor. Writing the field $\phi (\eta,x) = a(\eta)^{-1}\chi$, the equation of motion that follows from the action (\ref{m63}) is
\begin{equation}\label{m67}
\chi''- \nabla^2 \chi +\bigg( m^2a^2-{a''\over a}\bigg)\chi=0\,,
\end{equation}
where the prime denotes derivatives with respect to the conformal time $\eta$. We can see that the field $\chi$ obeys the same equation of motion as a massive scalar field in Minkowski space-time, but now with a time dependent {\em effective mass},
\begin{equation}\label{m68}
m^2_{eff}(\eta)\equiv m^2a^2-{a''\over a}\,.
\end{equation}
This time dependent  mass accounts for the interaction between the scalar field and the gravitational field.  The energy  of the field $\chi$ is not conserved (its action is explicitly time-dependent), and, more important, its quantization leads to particle creation at the expense of the classical gravitational background. However, for a FRW type Universe dominated by radiation one finds that $a'' = 0$. Therefore, for a massless scalar field, there is no particle production  since Eq. (\ref{m67}) reduces to same one of the Minkowski spacetime. This is the Parker result which was clearly deduced with basis on the general relativity \cite{park68}. In other words, by assuming the general relativity, the gravitational field of the Universe at early stages is unable to work as a pump supplying energy for massless scalar fields. 

After a Fourier expansion of Eq. (\ref{m67}), we are left with the mode equation
\begin{equation}\label{nv4}
\chi''_k+\omega_k^2(\eta)\chi_k=0\,,
\end{equation}
with $\omega_k^2(\eta)\equiv k^2+m^2_{eff}$. 
The solutions of the above equation give the positive- and negative-frequency modes. Following standard lines, the quantization can be carried out by imposing equal-time commutation relations for the scalar field $\chi$ and its canonically conjugate momentum $\pi\equiv\chi'$, namely $[\chi(x,\eta) \,, \pi(y,\eta)]=i\delta(x-y)$, and by implementing secondary quantization in the so-called Fock representation. After convenient Bogoliubov transformations, one obtains the transition amplitudes for the vacuum state and the associated spectrum of the produced particles in a non-stationary background \cite{books,mukh}. Many conceptual aspects related to possible ambiguities arising from the dependence of the particle number density with the reference frame and their  connections with moving detectors have also attracted attention in the literature (see, for instance, Refs. \cite {books,gribmama,detect}).

\section{Expanding Universe in $f(R)$ Theories}
\label{s3}

Let us briefly recall how the evolution  of the scale factor $a(\eta)$ is modified in the context of $f(R)$ theories. Following  Palatini's approach, where the metric and the connection are independent variables, the generalized Einstein equations for a self-gravitating fluid are given by \cite{gribmama,fR,allemandi,sotiriou,meng}
\begin{equation}\label{eq11}
f_{,R} \,R_{\mu\nu}(\Gamma)-\frac{1}{2}\,f(R)\,g_{\mu\nu}=8\pi G\,T_{\mu\nu}\;,
\end{equation}
where $f_{,R}\equiv df(R)/dR$, $R_{\mu\nu}(\Gamma)$ is the Ricci tensor of any independent torsionless connection $\Gamma$, $R\equiv g^{\alpha\beta}R_{\alpha\beta}(\Gamma)$ is the generalized Ricci scalar (which reduces to the ordinary Ricci scalar $\bar{R}$ only for some special cases) and $T_{\mu\nu}$ is the energy-momentum tensor.  The trace of the above equation reads
\begin{equation}\label{11a}
f_{,R}(R)R - 2f(R)=8\pi G \tau\,,
\end{equation}
where $\tau \equiv\, $tr$T_{\mu\nu}$, and for a perfect fluid it is $\tau=\rho-3p$. The above identity  controls solutions of Eq. (\ref{eq11}) in the sense that, given a $f$ function we can obtain solutions $R$ depending on $\tau$. In certain special cases these solutions can be obtained explicitly. In particular,  for a radiation or vacuum dominated universe the solution is simply $R=0$ or $R=const$, so that the solutions of the generalized Friedman equation can be explicitly obtained.  

Without loss of generality, we can choose $f(R)$ in the following form:
\begin{equation}\label{eq14}
f(R)=R + g(R)\,.
\end{equation}
In the general case of a medium with equation of state in the form, $p=\omega \rho$, one obtains a generalized FRW differential equation in terms of $R(\Gamma)$.  With the help of the trace identity as given by  Eq. (\ref{11a}), we can write the following equation for the expansion of the scale factor in the conformal time $\eta$:
\begin{equation}\label{eq15}
{a''\over a}+(\Delta -1)\bigg({a'\over a}\bigg)^2+a^2 G(\Delta,R)=0\,,
\end{equation}
with
\begin{equation}\label{eq16}
G(\Delta,R) = -{g\over 6}+{\Delta\over 6}(Rg_{,R}-g)+(\Delta-1)\bigg[H^2 g_{,R} +{R'\over a}g_{,RR}+ {1\over 4}\bigg( {R'\over a} \bigg)^2{g^2_{,RR}\over 1+g_{,R}}\bigg],\nonumber\\
\end{equation}
where $\Delta=(1+3\omega)/2$ and $H$ is the Hubble parameter. Note that if $g(R)=0$ then $G(\Delta,R)=0$ and the general relativistic FRW differential equation for the scale factor is recovered \cite{AssadLima}. It should also be recalled that only in the case $\tau=0$ (radiation), we have $R$ equal to the ordinary Ricci scalar $\bar{R}$.  As we shall see in the next section,  the solutions of Eq. (\ref{eq15}) in this case are easily obtained for some special  expressions of $g(R)$. 

\section{Particle Creation in $f(R)$ Gravity}

To begin with we consider the particle creation process in a radiation dominated universe, $\omega=1/3$ (or $\Delta=1$) for which  $R=\bar{R}$. In this case,  Eq. (\ref{eq15}) is simplified to: 
\begin{equation}\label{eq17}
{a''\over a}+a^2 G(1,\bar{R})=0\,,
\end{equation}
where
\begin{equation}\label{eq18}
G(1,\bar{R}) = -{g\over 6}+{1\over 6}(\bar{R}g_{,R} - g)\,,
\end{equation}
and $\bar{R}=-6a''/a^3$. Inserting Eq. (\ref{eq17}) into Eq. (\ref{m68}) we see that $G$ acts exactly as an effective mass, and, therefore, it works like a source of particle production even in the massless case. 

Let us now assume that the $g(R)$ function is a power law 
\begin{equation}
g(R)=\beta R^{n}\,, \hspace{1cm} (0<n<2)
\end{equation}
where $\beta$ is a dimensional constant. In this case Eq. (\ref{eq17}) takes the form
\begin{equation}
{a''\over a}+{1\over 6}[\beta(n-2)]^{-{1\over n-1}}a^2 =0\,,
\end{equation}

or, equivalently,
\begin{equation}
{a''}={1\over 6\alpha}a^3\,,\hspace{1cm}\alpha=[\beta(2-n)]^{{1\over n-1}}.
\end{equation}

For simplicity, in what follows it will be assumed that $\beta$ is positive.
An interesting particular solution of the above equation is
\begin{equation}
a(\eta)=-{\sqrt{12\alpha}\over \eta}\,, \hspace{1cm} -\infty <\eta<0\,,
\end{equation}
which has exactly the same form of a de Sitter expansion in general relativity \cite{mukh2}. 

Now, inserting the above result into Eqs. (\ref{m68}) and (\ref{nv4}), we find that the mode function satisfies
\begin{equation}\label{modeeq}
\chi_k''+\bigg[k^2-(1-6\alpha m^2){2\over \eta^2}\bigg]\chi_k=0\,\,.
\end{equation}
This is a second order differential equation whose independent solutions, $\chi_{1k}$ and $\chi_{2k}$, are given by 
\begin{equation}\label{modef1}
\chi_{1k}(\eta)=c_1\sqrt{|\eta|}J_\nu(k|\eta|)\,,
\end{equation}
\begin{equation}\label{modef2}
\chi_{2k}(\eta)=c_2\sqrt{|\eta|}Y_\nu(k|\eta|)\,,
\end{equation}
where $J_\nu,\, Y_\nu$ are the Bessel functions of first and second kind, respectively, and the mass dependent parameter, $\nu\equiv \sqrt{9/4-12\alpha m^2}$, defining the  order of the functions.  For all times, these modes must be normalized according to
\begin{equation}\label{norm}
W_k(\eta)\equiv \chi_{1k}(\eta)\chi'_{2k}(\eta)-\chi'_{1k}(\eta)\chi_{2k}(\eta)=-2i.
\end{equation}

It is also worth noticing that Eq. (\ref{modeeq}) has some interesting features. First, it has exactly the same form obeyed  by the  mode  functions in  the case of a de Sitter cosmology in the framework of general relativity. In other words, we start with a radiation dominated universe in a $f(R)$ theory and the evolution is described by the same equation of a pure de Sitter universe  in the linear case. Particle creation in the de Sitter universe in the context of GR has been studied by several authors \cite{deSitter,blm98,mijic}. Another interesting feature of Eq. (\ref{modeeq}) is that it does not depends on the parameters $\beta$ and $n$  in the masssless limit.

The Bogoliubov coefficients can be calculated and a straightforward calculation leads to the final expression for the number of particles created in the $k$ mode \cite{blm98,mijic}:
\begin{equation}\label{nk}
N_k(\eta)={1\over 2|\omega_k(\eta)|}|\chi'_{2k}(\eta)|^2+{|\omega_k(\eta)|\over 2}|\chi_{2k}(\eta)|^2 -{1\over 2}\,.
\end{equation}
This expression correctly reproduces the initial vacuum condition $N_k(\eta \to -\infty)=0$. 

The concentration of created particles is readily obtained by integrating over all the modes:
\begin{equation}\label{n}
n={1\over a(\eta)^3}\int N_k(\eta)d^3 k\,.
\end{equation}

Now it is easy to calculate the particle creation, at least for some specific cases.

\subsection{Massive Case}

In order to illustrate the phenomenon of particle creation in FRW type cosmological models inspired by $f(R)$ gravity, let us first consider some simple example for massive particles.  In the case  $\nu=1$, that is, for $m^2=5/48\alpha$, it is easy to check that the normalized solutions are:
\begin{equation}\label{modef1m}
\chi_{1k}(\eta)=\sqrt{{2\over k}}{J_1(k|\eta|)\over \sqrt{ Y_1(k|\eta|)J_0(k|\eta|) - Y_0(k|\eta|)J_1(k|\eta|)} }\,,
\end{equation}
\begin{equation}\label{modef2m}
\chi_{2k}(\eta)=i\sqrt{{2\over k}}{Y_1(k|\eta|)\over \sqrt{ Y_1(k|\eta|)J_0(k|\eta|) - Y_0(k|\eta|)J_1(k|\eta|)} }\,,
\end{equation}
and the corresponding  spectrum of created particles is readily obtained by inserting the above solutions into (\ref{nk}).  

In Fig. 1, we show the spectrum for two different wavenumbers. The spikes represent the points where the frequency vanishes and $N_k$ diverges.

\begin{figure}[t]
\begin{center}
\epsfig{file=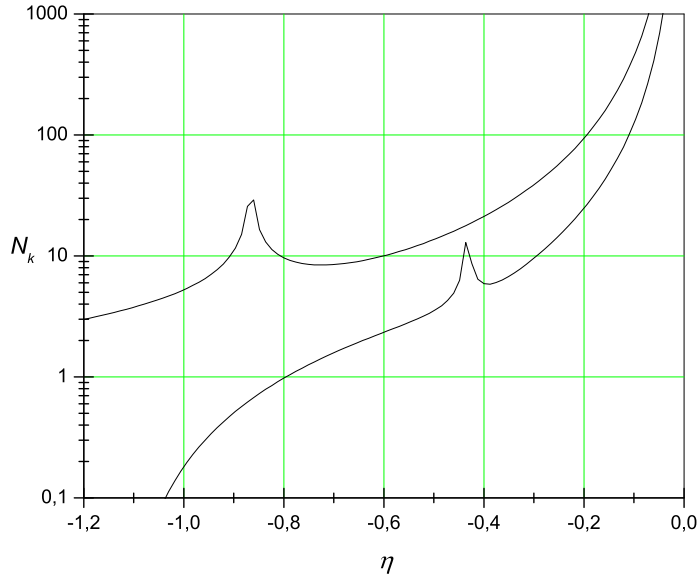, scale=1.1}
\end{center}
\caption{Massive Case. Spectrum of the created particles as function of the conformal time $\eta$ for two different values of the wavenumber $k$, from left to right, $k=1$ and $k=2$ respectively. The spikes represent the points where the frequency vanishes and $N_k$ diverges. }\label{fig1}
\end{figure}

Another interesting case is $m^2=1/6\alpha$ which is conformally related to flat space-time. The frequency vanishes for this choice of mass so that the modes are plane waves at all times, and, therefore, there is no particle production. The absence of matter creation in this $f(R)$ massive case is similar  to the classical result derived by Parker in general relativity for the massless case \cite{park68}. Implicitly, it also suggests the possibility of  massless particles production during the radiation dominated phase in $f(R)$ models. In order to show that, let us now calculate the spectrum of the created massless particles.

\subsection{Massless Case}

The main result of this paper is related to the creation of massless particles ($m=0$). In this case, the normalized mode functions (\ref{modef1}) and (\ref{modef2}) takes the following forms
\begin{equation}\label{modef1a}
\chi_{1k}(\eta)={1\over \sqrt{k}}\left(1+{i\over k\eta}\right)e^{ik\eta},
\end{equation}
\begin{equation}\label{modef2a}
\chi_{2k}(\eta)={1\over \sqrt{k}}\left(1-{i\over k\eta}\right)e^{-ik\eta}.
\end{equation}

\begin{figure}[t]
\begin{center}
\epsfig{file=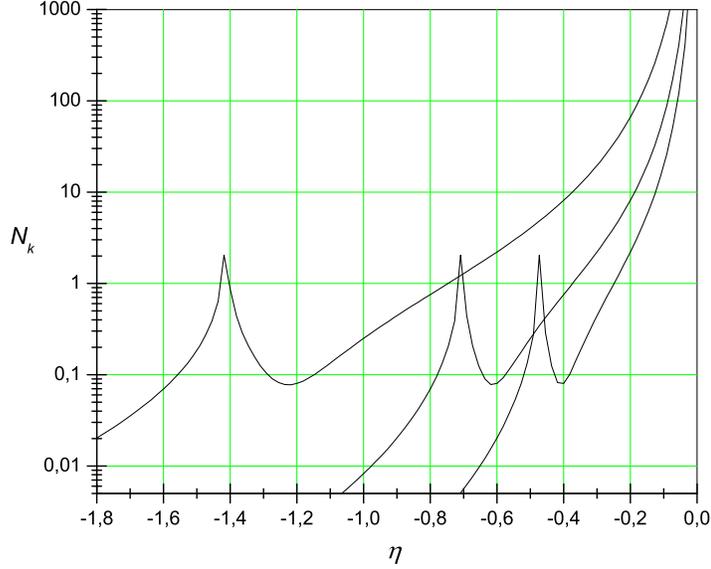, scale=1.1}
\end{center}
\caption{Massless Case. Spectrum of created particles as a function of the conformal time $\eta$ for different values of the wavenumber $k$, from left to the right, $k=1$, $k=2$ and $k=3$ respectively. The spikes represents the points where the frequency vanishes and $N_k$ diverges. }\label{fig2}
\end{figure}

Now, inserting the above expressions  into Eq. (\ref{nk}) we obtain the spectrum of created particles for each mode $k$:
\begin{eqnarray}
N_k(\eta)=-{1\over 2}-{1\over 4}|k^2\eta^2-2|^{1/2}\bigg({1\over k\eta}+ {1\over k^3\eta^3}\bigg)-{1\over 4|k^2\eta^2-2|^{1/2}}\bigg(k\eta-{1\over k\eta}+ {1\over k^3\eta^3}\bigg).\nonumber\\
\end{eqnarray}

In Fig. 2 we display the graphic of this function for some values of $k$. Apart from the spikes which occur due to the vanishing of the frequency, we also see that the number of created particles  grows  in the limit $\eta \to 0$.


\section{Concluding Remarks}
\label{s5}

Due to the  discovery of the accelerating Universe at low redshifts, an increasing attention has been paid to $f(R)$ gravity as a possibility to reduce the so-called cosmological dark sector.  In this kind of cosmology the Universe can be accelerating driven only by cold dark matter. 

In this work, assuming that the deviations from general relativity are described by a power law, we have addressed the problem of particle creation in $f(R)$ homogeneous and isotropic cosmologies dominated by a  fluid obeying a  $\omega$-type
equation of state. The mode equation in such background was derived and their general solution explicitly obtained. As an illustration, we have calculated the spectrum of created particles for a radiation dominated universe considering the cases of massive and massless particles.   

Unlike the Parker results which are based on the standard general relativity, we have also explicitly shown that massless particles can be produced during the radiation dominated phase in the framework of  $f(R)$ cosmologies.  This result is not valid only for the case of a scalar field as discussed here. Actually,  due to the new contributions to the FRW differential equation,  one may show that the positive- and negative-frequency parts  of the fields become mixed in an expanding $f(R)$ cosmology thereby leading to the creation of massless particles of nonzero spin even at early times.

\centerline{\bf Acknowledgments}
JASL is partially supported by CNPq and FAPESP No.
04/13668-0 and CHGB is supported by CNPq No 150429/2009-6 (Brazilian
Research Agency).


\begin{thebibliography}{0}

\bibitem{GWS83} R. Brout et. al. Ann. Phys. (N.Y.), {\bf 115}, 78 (1978); D. Atkatz and H. Pagels, Phys. Rev. D {\bf 25}, 2065 (1982); J. B. Hartle, in {\it The Very early Universe}, edited by G. Gibbons, S. W. Hawking, and S. Siklos (Cambridge UP, Cambridge, 1983); B. L. Hu and Pav\'on, Phys. Lett. B {\bf 180}, 329 (1986).

\bibitem{Zeld70} Ya. B. Zeldovich, JETP Lett. {\bf 12}, 307 (1970).

\bibitem{Murphy73} G. L. Murphy, Phys. Rev. {\bf D48}, 4231 (1973); Z. Klimek, Acta Cosmologica {\bf 3}, 49 (1975).

\bibitem{Hu82} B. L. Hu, Phys. Lett. {\bf A90}, 375 (1982).
Adv. Astrop. {\bf 1}, 23 (1983). 

\bibitem{Barrow86} J. D. Barrow, Phys. Lett
{\bf B180}, 335 (1986); M. Morikawa and M. Sasaki, Phys. Lett. B {\bf 165}, 59 (1985); T. Padmanabhan and S. M. Chitre, Phys. Lett. A {\bf 120} 433 (1987);  J. A. S. Lima, R. Portugal and I. Waga, Phys. Rev. D {\bf 37}, 2755 (1988). 

\bibitem{Prigogine} I. Prigogine, J. Geheniau, E. Gunzig, P. Nardone,
 Gen. Rel. Grav. {\bf 21}, 767 (1989).

\bibitem{CLW} M. O. Calv\~{a}o, J. A. S. Lima, I. Waga, Phys. Lett.
{\bf A162}, 223 (1992). See also, J. A. S. Lima, M. O. Calv\~{a}o, I. Waga, ``Cosmology,
Thermodynamics and Matter Creation'', {\it Frontier Physics, Essays in
Honor of Jayme Tiomno}, Edited by , (World Scientific, Singapore, 1991). 

\bibitem{AGL} J. A. S. Lima, A. S. M. Germano, Phys. Lett. {\bf A170},
373 (1992); W. Zimdahl and D. Pav\'{o}n, Mon. Not. R. Astr.
Soc. {\bf 266}, 872 (1994); W. Zimdahl and D. Pav\'{o}n, Gen. Rel. Grav. {\bf 26}, 1259  (1994); L. R. W. Abramo and J. A. S. Lima, Class. Quant. Grav. {\bf 13}, 2953 (1996).

\bibitem{LGA}J. Gariel, G. Le Denmat, Phys. Lett. {\bf 
A200}, 11 (1995); J. A. S. Lima, A. S. M. Germano and L. R. W. Abramo, Phys. Rev. {\bf D53}, 4287 (1996).
\bibitem{zimd} J. A. S. Lima and J. S. Alcaniz,  Astron. Astrophys. {\bf 348}, 1 (1999), [astro-ph/9902337]; Zimdhal, D. J. Schwarz, A. B. Balakin  and D. Pav\'on, Phys. Rev. D {\bf 64}, 063501 (2001). 

\bibitem{LSS08} J. A. S. Lima, F. E. Silva and R. C. Santos, Class. Quant. Grav. {\bf 25}, 205006 (2008), arXiv:0807.3379 [astro-ph]; G. Steigman, R. C. Santos and J. A. S. Lima, JCAP {\bf 06}033 (2009), arXiv:0812.3912 [astro-ph].  

\bibitem{park68} L. Parker, Phys. Rev. Lett. {\bf 21}, 562 (1968).

\bibitem{park69} L. Parker, Phys. Rev. {\bf 183}, 1057 (1969); Phys. Rev. {\bf D 3}, 346 (1971); Phys. Rev. Lett. {\bf 28}, 705 (1972); Phys. Rev. {\bf D 7}, 976 (1973).

\bibitem{books} N. D. Birrell and P. C. W. Davies, {\em Quantum Fields in Curved 
Space} (Cambridge University Press, Cambridge, 1982);
S. A. Fulling, {\em Aspects of Quantum Field Theory in Curved 
Spacetime} (Cambridge University Press, Cambridge, 1989); A. A. Grib, S. G. Mamayev and V. M. Mostepanenko, {\em Vaccum Quantum effects in Strong Fields} (Friedmann Laboratory Publishing, St. Petesburg, 1994).

\bibitem{mukh} V. F. Mukhanov and S. Winitzki, {\em Introduction to Quantum Effects in Gravity}, (Cambridge University Press, Cambridge, 2007).

\bibitem{partcrea} Ya. B Zel'dovich, Pisma Zh. Eksp. Teor. Fiz. {\bf 12}, 443 (1970), (English transl. JETP {\bf 12}, 307 (1970)); A. A. Grib, S. G. Mamayev and V. M. Mostepanenko, Gen. Rel. Grav., {\bf 7}, 535 (1975); A. A. Grib and Yu. V. Pavlov, [gr-qc/0505140]; Grav. Cosmol. {\bf 11}, 119 (2005); Grav. Cosmol. {\bf 12}, 159 (2006). 

\bibitem{grav} L. P. Grishchuk, Class. Quant. Grav. {\bf 10}, 2449 (1993); M. R. G. Maia, Phys. Rev. D {\bf 48}, 647 (1993); M. R. G. Maia and J. D. Barrow, Phys. Rev. {\bf D 50}, 6262 (1994); M. R. G. Maia and J. A. S. Lima, Phys. Rev. {\bf D 54}, 6111 (1996).

\bibitem{gribmama} A. A. Grib and S. G. Mamayev, Yad. Fiz. {\bf 10}, 1276 (1969)[English transl.: Sov. J. Nucl. Phys. {\bf 10}, 722 (1970)]. See also Yu. V. Pavlov, Theor. Math. Phys. {\bf 126}, 92 (2001), [gr-qc/0012082].

\bibitem{detect} M. I. Shirokov, Sov. J. Nucl. Phys. (USA), {\bf 7}, 411 (1968); A. A. Grib and S. G. Mamayev, Sov. J. Nucl. Phys. (USA), {\bf 14}, 450 (1972); J. R. Letaw and J. D. Pfautsch, J. Math. Phys. {\bf 23}, 425 (1982); W. Junker and E. Schrobe, Ann. Inst. Henry Poincar\'e {\bf 3}, 1113 (2002); L. C. Crispino, A. Higuchi and G. E. A. Matsas, Rev. Mod. Phys. {\bf 80}, 787 (2009).

\bibitem{fR} D. N. Vollick, Phys. Rev. {\bf D 68}, 063510 (2003); 
M. Amarzguioui, O. Elgaray, D. F. Mota and T. Multamaki, Astron. Astrophys. {\bf 454}, 707, (2006); S. Nojiri and S. D. Odintsov, Phys. Lett. {\bf B 659}, 821, (2008).


\bibitem{allemandi} G. Allemandi, A. Borowiec, M. Francaviglia and S. D. Odintsov, Phys. Rev. {\bf D 72}, 063505 (2005); L. Amendola, D. Polarski and S. Tsujikawa, Phys. Rev. Lett. {\bf 98}, 131302 (2007); J. Santos, J. S. Alcaniz, F. C. Carvalho and  N. Pires, Phys. Lett. {\bf B 669}, 14 (2008); J. Santos and  M. J. Reboucas, Phys. Rev. D {\bf 80}, 063009 (2009). For a review see, T. P. Sotiriou and V. Faraoni, arXiv:0805.1726v2, [gr-qc], Rev. Mod. Phys. (2009). To appear.

\bibitem{gaspe} M. Gasperine and G. Veneziano, Phys. Lett. {\bf B 277}, 256 (1992).

\bibitem{vilk} G. A. Vilkovisky, Class. Quant. Grav. {\bf 9}, 895 (1992).

\bibitem{staro} A. A. Starobinsky, Phys. Lett. {\bf B 91}, 99 (1980).

\bibitem{dolgov} A. D. Dolgov and M. Kawasaki, Phys. Lett. {\bf B 573}, 1 (2003); T. Chiba, Phys. Lett. {\bf B 575}, 1 (2003).

\bibitem{cembranos} J. A. R. Cembranos, Phys. Rev. {\bf D 73}, 064029 (2006); T. P. Sotiriou, Class. Quant. Grav. {\bf 23}, 1253 (2006). 

\bibitem{sotiriou} T. P. Sotiriou, Phys. Rev. {\bf D 73} 063515 (2006). 

\bibitem{meng} X. Meng and P. Wang,  Class. Quant. Grav. {\bf 21}, 2029 (2004).

\bibitem{waga09} V. Miranda, S. E. Jor\'as, I. Waga and M. Quartin, Phys. Rev. Lett. {\bf 102}, 221101 (2009).

\bibitem{AssadLima} M. J. D. Assad and J. A. S. Lima, Gen. Rel. Grav. {\bf 20}, 527 (1988); J. A. S. Lima, Am. J. Phys. {\bf 69}, 1245 (2001).

\bibitem{mukh2} V. Mukhanov, {\it Physical Foundations of Cosmology}, Cambridge University Press, (2005).

\bibitem{blm98} M. Bordag, J. Lindig and V. M. Mostepanenko, Class. Quantum Grav. {\bf 15} 581 (1998).

\bibitem{deSitter} E. Mottola, Phys. Rev. {\bf D 31}, 754 (1985);  C. Molina-Par\'is, Int. J. Th. Phys. {\bf 38}, 1273 (1999).
 
\bibitem{mijic} M. Mijic, Phys. Rev. {\bf D 57}, 2138 (1998); S. Biswas and I. Chowdhury, Int. J. Mod. Phys. {\bf D 15}, 937 (2006).

\end{thebibliography}
\end{document}